\documentclass[12pt]{iopart}
\usepackage{graphicx,epsfig}
\usepackage{color}
\usepackage{cite}

\begin{document}

\title{Coevolution of strategies and update rules in the Prisoner's Dilemma game on complex networks}

\author{Alessio Cardillo}

\address{Dipartimento di Fisica e Astronomia, Universit\`a di Catania
  and INFN Sezione di Catania, Catania I-95123, Italy}

\author{Jes\'us G{\'o}mez-Garde\~{n}es}

\address{Departamento de Matem\'atica Aplicada, Universidad Rey Juan
  Carlos, M\'ostoles (Madrid) E-28933, Spain}

\address{Institute for Biocomputation and Physics of Complex Systems
  (BIFI), University of Zaragoza, Zaragoza 50009, Spain}

\author{Daniele Vilone}

\address{Grupo Interdisciplinar de Sistemas Complejos (GISC),
  Departamento de Matem\'aticas, Universidad Carlos III de Madrid,
  28911 Legan\'es, Spain}
  
\author{Angel S\'anchez}

\address{Institute for Biocomputation and Physics of Complex Systems
  (BIFI), University of Zaragoza, Zaragoza 50009, Spain}

\address{Grupo Interdisciplinar de Sistemas Complejos (GISC),
  Departamento de Matem\'aticas, Universidad Carlos III de Madrid,
  28911 Legan\'es, Spain}
  
\address{Instituto de Ciencias Matem\'aticas CSIC-UAM-UC3M-UCM, 28049
  Madrid, Spain}

\date{\today}

\begin{abstract}
In this work we study a weak Prisoner's Dilemma game in which both
strategies and update rules are subjected to evolutionary
pressure. Interactions among agents are specified by complex
topologies, and we consider both homogeneous and heterogeneous
situations. We consider deterministic and stochastic update rules for
the strategies, which in turn may consider single links or full
context when selecting agents to copy from. Our results indicate that
the co-evolutionary process preserves heterogeneous networks as a
suitable framework for the emergence of cooperation. Furthermore, on
those networks, the update rule leading to a larger fraction, which we
call replicator dynamics, is selected during co-evolution. On
homogeneous networks we observe that even if replicator dynamics turns
out again to be the selected update rule, the cooperation level is
larger than on a fixed update rule framework. We conclude that for a
variety of topologies, the fact that the dynamics coevolves with the
strategies leads in general to more cooperation in the weak Prisoner's
Dilemma game.
\end{abstract}


\maketitle

\section{Introduction}

Evolutionary game theory on graphs or networks has attracted a lot of
interest among physicists in the last decade
\cite{szabo:2007,roca:2009a}, both because of the new phenomena that
such a non-hamiltonian dynamics \cite{hofbauer:1998} gives rise to and
because of its very many important applications
\cite{nowak:2006a,gintis:2009}. Prominent among these applications is
the understanding of the emergence of cooperation among bacteria,
animals and humans \cite{pennisi:2009}. In this context, evolutionary
game theory on graphs is relevant due to the fact that populations
interact with a reduced, fixed subset of the population has been
proposed as one of the mechanisms leading to cooperation among
unrelated individuals \cite{nowak:2006b}. Indeed, the existence of a
network of interactions may allow the assortment of cooperators in
clusters \cite{gomezgardenes:2007,fletcher:2009}, that can outperform
the defectors left on their boundaries.  Unfortunately, extensive
research has shown that this proposal is not free from difficulties,
in particular due to the lack of universality of the evolutionary
outcomes when changing the network or the dynamics
\cite{roca:2009a,roca:2008}. In addition, recent experimental tests of
the effect of networks on the evolution of cooperation
\cite{traulsen:2010,grujic:2010} have led to new questions in so far
as the experimental results do not match any of the models available
in the literature.

In this work, we address the issue of the lack of universality and the
problems it poses for applications to real world problems through the
idea of co-evolution.  The rationale behind this approach is simple:
if there are many possibilities regarding networks or dynamics and no
{\em a priori} reasons to favor one over another, one can take a step
beyond in evolutionary thought by letting a selection process act on
those features: The types of networks or dynamics that are not
selected along the process should not be considered then as
ingredients of applicable models. This co-evolutionary perspective has
led to a number of interesting results in the last few years (see,
e.g., \cite{gross:2008,perc:2010} for reviews), starting from the
pioneering work by Ebel and Bornholdt \cite{ebel:2002a} on
co-evolutionary games on networks. Subsequently, a fruitful line of
work has developed by allowing players to rewire their links (e.g.,
\cite{zimmermann:2004,pacheco:2006,pestelacci:2008,szolnoki:2009,szolnoki:2009a,szolnoki:2009b,wu:2010})
or even forming the network by adding new players with attachment
criteria depending on the payoffs
\cite{poncela:2008,poncela:2009a,poncela:2009b}. We here take a much
less explored avenue, namely the simultaneous evolution of the
players' strategies and the way they update them. This idea, first
introduced in an economic context by Kirchkamp \cite{kirchkamp:1999}
has been recently considered in biological and physical contexts
\cite{moyano:2009,szabo:2009,szolnoki:2009c}, proving itself promising
as an alternative way to understand the phenomenon of the emergence of
cooperation.

Here we largely expand on the work by Moyano and S\'anchez
\cite{moyano:2009}, where only the case of a square lattice was
considered. For that specific example, it was shown that evolutionary
competition between update rules gives rise to higher cooperation
levels than in a well-mixed context (complete graph) or when the
update rules are fixed and cannot evolve. We now aim to address this
issue in a much more general setup as regards the network of
interactions, by considering a family of complex networks proposed in
\cite{gomezgardenes:2006} as the structure of the population. This
will allow us to go from complex but homogeneous networks to scale
free ones, thus providing insight into what effects arise because of
the specific degree distribution or because of the complexity of the
network. As we will see below, a wide variety of effects and results
will emerge as a product of this co-evolutionary process in this class
of networks. In order to present our findings, Section \ref{sec:2}
introduces our model of the co-evolutionary process and recalls the
main features of the one-parameter family of networks we will use as
substrate. Section \ref{sec:3} collects our results, organized in
terms of the competition between pairs of update rules. Finally,
Sec.\ \ref{sec:4} summarizes our conclusions and discusses their
implications.

\section{The model}
\label{sec:2}

Following the scheme introduced for the first time in
\cite{moyano:2009}, our model is formulated as follows: We consider a
set of agents with two associated features: a strategy and an update
rule. The strategy can take two possible values, namely cooperation
(C) and defection (D), whereas for the update rule agents can take one
of the following three options: Replicator (REP), Moran-like (MOR) and
Unconditional Imitation (UI) rules, which we will define precisely
below.  These two features of individuals coevolve simultaneously
according to the evolutionary dynamics specified below thus yielding
an interesting feedback-loop between the game strategies and learning
rules of the population. The co-evolution process is specified by
stipulating that when an individual, applying her update rule, decides
to copy the strategy of a neighboring agent, she copies this agent's
update rule as well, i.e., the update process involves simultaneously
the two features of the agent. While this is clearly a more
biologically inspired approach, it is also possible to think of
economic contexts in which not only decisions are imitated but also
the way to arrive to those decisions can be copied as well.

As usual in the studies on evolutionary dynamics on complex networks,
every player is represented by a node of the network. Agents thus
arranged interact through a game only with those players that are
directly connected to them as dictated by the underlying network.
Mathematically, the network is encoded by its adjacency matrix ${\bf
  A}$, whose elements are $A_{ij}=A_{ji}=1$ if nodes $i$ and $j$ are
connected and $A_{ij}=A_{ji}=0$ otherwise. Therefore, given a network
substrate the dynamics of each individual is driven by the local
constrains imposed by the topology. In order to study the effects of
topology on the evolution of the population we will study a family of
different network topologies generated by the model introduced by
G\'omez-Garde\~nes and Moreno in \cite{gomezgardenes:2006}.  This
family of networks depends on one parameter of the model, $\alpha \in
[0, 1]$, that measures the degree of heterogeneity of the degree
distribution, $P(k)$, {i.e.} the probability of finding nodes with
degree $k$.  Without going into full details of the model, that can be
found in the original reference \cite{gomezgardenes:2006}, we find it
convenient for the reader to briefly summarize its
definition. Networks are generated starting from a fully connected
graph with $m_0$ nodes and a set $\cal U$ of $N-m_0$ unconnected
nodes.
At each iteration of the growing
  process a new node in the set $\cal U$ is chosen to create $m$ new
  links. Each of these $m$ links are formed as follows: With
  probability $\alpha$ the link connects to any of the other $N-1$
  nodes in the network with uniform probability; with probability
  $(1-\alpha)$ it links to a node following a preferential attachment
  strategy {\em \'a la} Barab\'asi and Albert (BA)
  \cite{barabasi:1999}. This procedure is repeated until all the
  $N-m_0$ nodes in $\cal U$ have been chosen.  
Obviously, when $\alpha=0$ the model generates BA scale-free networks
with $P(k)\sim k^{-3}$. On the other hand, when $\alpha=1$ the
networks generated have a Poisson degree distribution as in
Erd\H{o}s-R\'enyi (ER) graphs \cite{erdos:1959}. For intermediate values
of $\alpha$ one goes from the large heterogeneity of scale-free
networks for $\alpha=0$ to the homogeneous distributions found for
$\alpha=1$. It is worth stressing that the model preserves the total
number of links, $L$, and nodes, $N$, for a proper comparison between
different values of $\alpha$.  In the following we will focus on the
cases $\alpha=0$ (BA), $0.5$ (intermediate) and $1.0$ (ER).

Having defined the network that specifies the agents a given one will
interact with, we now need to define how that interaction takes
place. Connected pair of nodes play the so-called weak Prisoner's
Dilemma, introduced in the pioneering paper on games on graphs by
Nowak and May \cite{nowak:1992}:
\begin{equation} 
\label{eq:pdnowakmay}
\begin{array}{ccc}
 & \mbox{ }\, {\rm C} & \!\!\!\!\!\! {\rm D} \\
\begin{array}{c} {\rm C} \\ {\rm D} \end{array} & \left(\begin{array}{c} 1 \\ b
\end{array}\right. & \left.\begin{array}{c} 0 \\ 0
\end{array}\right). \end{array} 
\end{equation}
With this choice for the payoff matrix, when two agents cooperate with
each other they both receive a payoff $1$. On the contrary, if one is
a defector and the other plays as cooperator the former receives $b>1$
and the latter receives nothing. Finally, if both nodes are defectors
they do not earn anything.  This setting is referred to as a weak
Prisoner's Dilemma game because there is no risk in cooperating as in
the standard Prisoner's Dilemma, in which a cooperator obtains a
negative payoff (i.e., she is punished) when facing a
defector. Therefore, in the standard Prisoner's Dilemma cooperating
involves the risk of losing payoff, whereas in its weak version that
is not the case, being the temptation to defect, $b$, the only
obstacle to cooperation.

\begin{figure}[!t]
\flushright
\epsfig{file=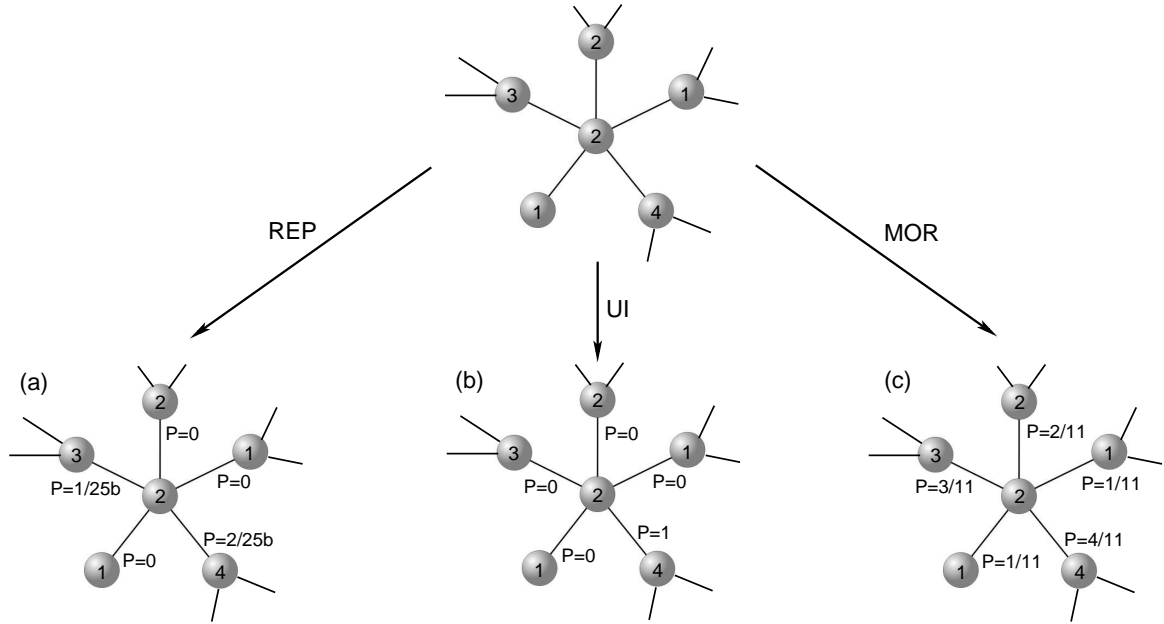,width=0.98\columnwidth,angle=0,clip=1}
\caption{Example of update dynamics according to the three rules we use
  in this work. On the top we represent a small portion of the network
  consisting of a central agent with five neighbours. The numbers inside the
  circles are the hypothetical payoffs of each player. The three
  pictures in the bottom represent the different possibilities for the
  central player to update her strategy. The probability to copy a
  given neighbour is indicated by its side. First, case {\bf (a)}
  illustrates a central player with REP update. She will imitate
  only neighbours doing better than herself (those with nonzero
  probability) with probability proportional to the 
  corresponding payoff difference
  with her own (that with the largest probability is the most likely
  to be imitated); in the second picture {\bf (b)}, explaining the UI
  dynamics, the central agent will imitate definitely her best
  neighbour, {\em i.e.}, the one with the largest payoff (provided it
  is larger than her own); finally, in the MOR update rule {\bf (c)},
  all the neighbours can be in principle imitated (provided they do
  not receive the absolute minimum payoff), with a probability
  proportional to their respective payoffs.}
\label{nueva}
\end{figure}

The last ingredient needed to specify our model, and the main focus of
this study, is the evolutionary dynamics. At a given time step, every
player plays the game with all her neighbors using the same strategy
in all pairings, as usual in evolutionary game theory on graphs. Once
all the games are played, each agent collects the total payoff
obtained from playing with her neighbors, $f_{i}$. Subsequently,
agents have to decide what strategy they will use in the next round,
which they do by means of some update rule. For the purposes of the
present paper, we will consider only update rules based on the payoff,
those being more relevant to biological applications (where ``payoff''
is interpreted as ``fitness'' or reproductive capacity). Among those,
we will consider three of the most often used rules
\cite{szabo:2007,roca:2009a} (see Fig.\ \ref{nueva} for a sketch of
the different rules to help grasp the differences between them):
\begin{itemize}
\item {\em Replicator Dynamics}\/ (REP)
  \cite{helbing:1992,schlag:1998}: Agent $i$ chooses one neighbor at
  random, say agent $j$, and compare their respective payoffs. If
  $f_{j}>f_{i}$, agent $i$ will adopt the state (i.e., the strategy
  and the update rule) of agent $j$ with probability:
\begin{equation}
\Pi=\frac{f_{j}-f_{i}}{b\max{(k_{i},k_{j})}}\;,
\end{equation}
where the denominator is chosen to ensure proper normalization of the
probability.  In case $f_{i}>f_{j}$, agent $i$ will keep the same
dynamical state in the following generation.  A remark is in order
there to indicate that REP is one of an infinite number of rules that,
when the number of agents is taken to infinity on a complete graph,
converges to the famous replicator equation \cite{hofbauer:1998},
hence the name we have used. Other authors prefer to call it
``proportional imitation'' to emphasize the fact that other rules have
the same limit, but we have kept the name REP because as it is found
in most textbooks (e.g., \cite{gintis:2009}).

\item {\em Unconditional Imitation}\/ (UI) \cite{nowak:1992}: Agent
  $i$ compares its payoff with her neighbor with the largest payoff,
  say agent $j$. If $f_{j}>f_{i}$ agent $i$ will copy the two features
  of agent $j$ for the next round of the game. Otherwise, agent $i$
  will remain unchanged.

\item {\em Moran rule}\/ (MOR) \cite{moran:1962,roca:2008}: Agent $i$
  chooses one of its neighbors proportionally to its payoff, being the
  probability of choosing agent $j$ given by:
\begin{equation}
P_{j}=\frac{A_{ij}f_{j}}{\sum_{l}A_{il}f_{l}}\;. 
\end{equation}
Subsequently, agent $i$ changes its state to the one of the chosen neighbor.
\end{itemize}

We stress that, as stated above, and at variance with traditional
evolutionary dynamics, the above three update mechanisms refer to the
dynamical state of the nodes, here defined by both the strategy and
the update rule. The three rules we have chosen allow us to consider
different aspects of the dynamics. Indeed, REP is stochastic and
link-focused (by this we mean that the update is carried out after
just looking at one of links of the agent); UI is deterministic and
context-focused (the update depends simultaneously on all the
neighbors of the considered agent), and MOR is stochastic and
context-focused. A deterministic, link-focused rule would proceed to
choose a link at random and copy the agent in case that her payoff is
larger. However, one can see that this is basically the same as REP
with a different probability normalization, and that would simply
amount to changing the time scale of the simulation. On the other
hand, only the first two rules are monotonous, in the sense that
evolution proceeds always in the direction of increasing fitness, and
players do not make mistakes. This is not the case of MOR, whereas the
strategy for updating is chosen among all neighbors with positive
probability, irrespective of whether they do better or not.  This is
not the only non-monotonous rule, and in fact one can introduce the
possibility of such errors in several manners, the most popular of
them being the so-called Fermi rule \cite{szabo:1998} (see also
\cite{altrock:2009} for a generalization that interpolates between the
standard Fermi rule and UI). We have not included this rule here to
avoid a proliferation of parameters (Fermi rule depends on a
temperature-like parameter that controls the number of mistakes) and
of possible pairs of strategies to compete, but it is clear that this
is another interesting example of non-monotonous, payoff-dependent
update rule, and we will comment on it in the conclusions.

In all cases, simulations proceed as follows: We begin with an initial
configuration composed equally of cooperators and defectors, randomly
assigning the strategy of each of the $N$ individuals with equal
probability for both. For the update rule, the second dynamical
feature, it is also randomly distributed. In order to understand the
mechanisms at work in the competition of rules, we consider only the
coexistence of two update rules in the population, distributed
uniformly among cooperators and defectors.  Thus, if $A$ and $B$ are
the two updates rules chosen for the simulation, and $x$ is the
initial fraction of individuals having $A$ rule then the initial
setting is composed of $xN/2$ individuals with initial state ($C,A$),
$(1-x)N/2$ with ($C,B$), $xN/2$ with ($D$,$A$) and $(1-x)N/2$ with
($D$,$B$). From this initial configuration, the simulation proceeds
round by round, with the payoffs being reset to zero after every round
in agreement with the general procedure (i.e., we do not accumulate
payoffs).  Simulations are run until a stationary state is reached,
which we verified it is achieved after 4000 rounds. We then measured
the magnitudes of interest averaging over another 100 rounds, and the
whole process was repeated for 100 realizations for every set of
parameters. We varied the value of $b$ from 1 to 2 in steps of 0.025,
and the frequency of one of the rules in the initial condition from 0
to 1, also in steps of 0.025. We used networks with 5000 nodes and
mean degree $\langle k \rangle=6$ for $\alpha=0, 0.5$ and 1.  Finally,
it is worth mentioning that individuals update their dynamical states
in parallel. This is the commonly adopted choice in simulations of
evolutionary game theory and, although some differences have been
reported in specific cases with sequential dynamics
\cite{huberman:1993}, changes are in general limited to a narrow range
of parameters \cite{nowak:1994,roca:2009a,roca:2008}, at least for
homogeneous networks (but see Fig.\ 4b in \cite{roca:2009b} for a
specific example of large differences in scale-free networks under UI
dynamics).

\section{Results}
\label{sec:3}

In this section, we present the results obtained when pairs of different update rules
are subject to the selection pressure arising from the co-evolutionary 
process. We have chosen to focus on two main
quantities: The average level of cooperation in the asymptotic regime,
$\langle c\rangle$, and the final composition of the population
in terms of the update rule. As we will see, these two magnitudes are 
already enough to understand all the mechanisms at work during the 
co-evolution. We have also looked at the dynamics of individual realizations
to ensure that our interpretation of the results is correct. In what follows, 
we discuss our findings
for the three possible competitions between pairs of update rules:
Moran versus Replicators, Moran versus Imitators and Replicators versus
Imitators.

\subsection{Moran versus Replicators}

\begin{figure}[t!]
\epsfig{file=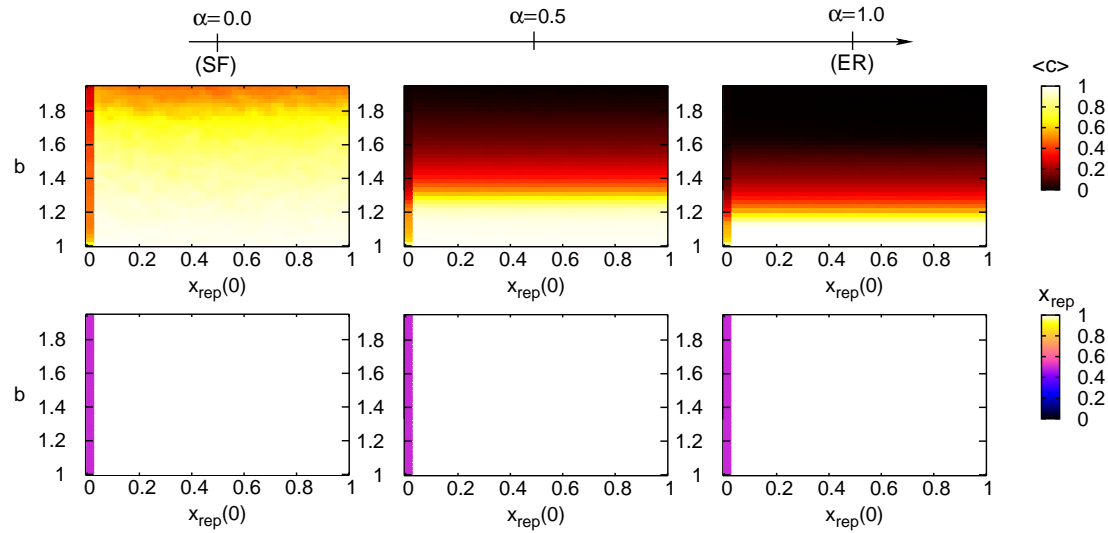,width=0.98\columnwidth,angle=0,clip=1}
\caption{{\em Moran versus Replicator}. The panels in the top denote
  the fraction of cooperators, $\langle c\rangle$, in the asymptotic
  regime in BA networks (left) networks with a moderate degree of
  heterogeneity, denoted as $\alpha=0.5$, (middle) and ER graphs
  (right). The panels in the bottom show the final fraction of
  replicators, $x_{rep}$ for the same cases as above. In all the
  panels both the degree of cooperation and the fraction of
  replicators are plotted as a function of the temptation $b$ and the
  initial fraction of replicators $x_{rep}(0)$.}
\label{morvsrep}
\end{figure}

To begin with, we deal with the two stochastic update rules, MOR and
REP. The results are shown in the panels of Figure \ref{morvsrep}. The
top panels show the degree of cooperation in the asymptotic regime,
$\langle c\rangle$, while the bottom panels show the final composition
of the population in terms of the fraction of replicator players,
$x_{rep}$. Both $\langle c\rangle$ and $x_{rep}$ are shown as a
function of the temptation to defect $b$ and the initial fraction of
replicators $x_{rep}(0)$ and for BA networks (left panels), networks
with moderate heterogeneity, corresponding to $\alpha=0.5$ in the
model introduced in \cite{gomezgardenes:2006}, (panels in the middle)
and ER graphs (right panels).

The bottom panels give the clue to interpret the processes that are
taking place in this first case.  REP players completely replace MOR
players regardless of the topology of the substrate network and the
values of $b$ and $x_{rep}(0)$ (provided
$x_{rep}(0)>0$). Consequently, the level of cooperation (top panels)
only depends on the ability of REP agents to cooperate. This ability,
that has been extensively studied (see \cite{szabo:2007,roca:2009a}
for detailed reviews) depends on both the temptation $b$ and the
substrate topology. Generally speaking, homogeneous networks with the
replicator rule support nonzero cooperation levels for small $b$, up
to $b\simeq 1.2$, whereas heterogeneous networks are much better
suited for cooperation under this dynamics
\cite{santos:2005,gomezgardenes:2007}.  The results for the
cooperation level, shown in the bottom panel, make it clear that we
recover the results of those previous studies as cooperation is
strongly enhanced when the heterogeneity of the network
increases. Thus, the MOR rule is basically irrelevant in the presence
of REP, a conclusion that agrees with that found in \cite{moyano:2009}
for the special case of square lattices.

\subsection{Moran versus Unconditional Imitation}

\begin{figure}[!t]
\epsfig{file=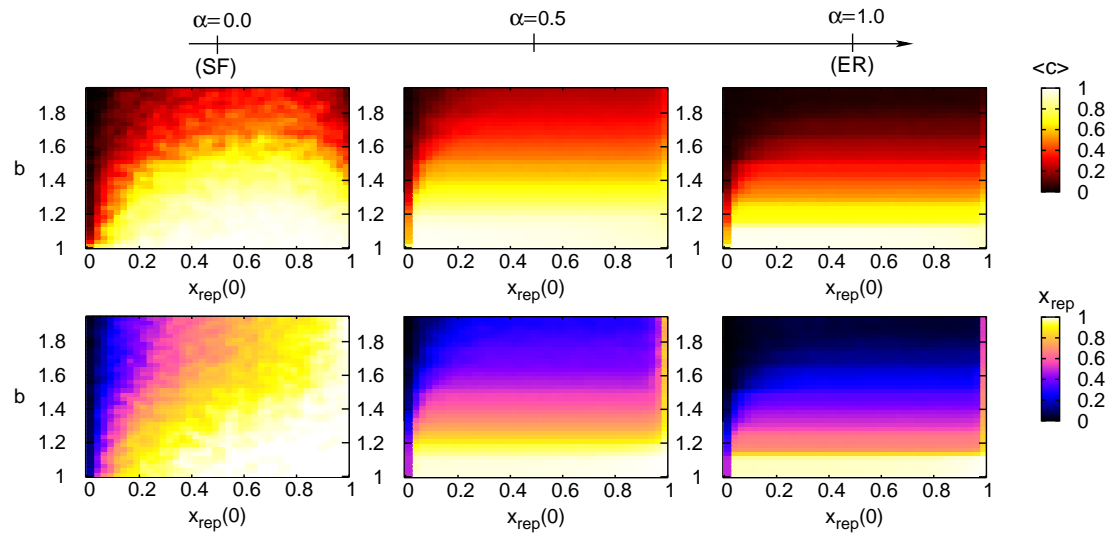,width=0.98\columnwidth,angle=0,clip=1}
\caption{{\em Moran versus Unconditional Imitation}. The panels in the
  top denote the fraction of cooperators, $\langle c\rangle$, in the
  asymptotic regime in BA networks (left) networks with a moderate
  degree of heterogeneity, denoted as $\alpha=0.5$, (middle) and ER
  graphs (right). The panels in the bottom show the final fraction of
  imitators, $x_{ui}$ for the same cases as above. In all the panels
  both the degree of cooperation and the fraction of imitators are
  plotted as a function of the temptation $b$ and the initial fraction
  of imitators $x_{ui}(0)$.}
\label{morvsui}
\end{figure}

Having seen that MOR can not survive in the presence of REP in any of
the networks considered, it is important to consider the case of UI vs
MOR, a deterministic rule versus a stochastic, non-monotonous one.  If
MOR is also suppressed by UI, we can then conclude that it is a very
fragile rule and its relevance will be very limited. In
Fig.\ \ref{morvsui} we plot the final density of cooperators (top
panels) and the final fraction of imitators (bottom panels) for the
three types of networks used in this study, BA networks (left panels),
moderate heterogeneous networks (center panels), and ER graphs (right
panels). For this particular competition, we observe that the
competition between these two rules is closely associated to the
emergence of cooperation: UI suppreses MOR only when cooperation is
the asymptotic result, and both quantities $\langle c\rangle$ and
$x_{ui}$ show very similar values.  In particular, for ER networks UI
players turn out to be responsible for all the cooperative behavior
observed and the evolutionary dynamics always ends up in a partition
of the population into MOR defectors and UI cooperators. It is
interesting to note that in this scenario MOR defectors are more
disruptive of the global cooperation level than in the case of a
regular square lattice, where cooperation survived until larger values
of the temptation $b$ \cite{moyano:2009}. We note that, for this weak
PD, when UI is the only update rule it leads to a higher level of
cooperative behavior on the entire range of $b$. This is very likely
due to the fact that the ER networks we are studying, having an
average degree of $\langle k\rangle=6$, are already close to a
well-mixed population, in which UI leads to full defection in one
step. Therefore, the net effect of having interaction between MOR and
UI players is that the cooperation level is reduced and the already
mentioned separation of strategists of two kinds.
  
When we move to more heterogeneous networks, the situation begins to
change.  As shown in the panels of both BA and $\alpha=0.5$ networks,
some Imitator-Defectors start to appear when the underlying topology
of interactions becomes more heterogeneous: Note that the level of
cooperation, top panels, is lower than the fraction of imitators, in
particular for large $b$ and for large initial fraction of imitator
players $x_{ui}(0)$. For the specific case of the BA network, the
invasion of imitators is complete in this limit, appearing when
$x_{ui}(0)>0.6$.  In the intermediate case, when $\alpha=0.5$, there
is a noticeable fraction of imitator defectors, as we already
mentioned, but the invasion of imitators is not complete.  In this
regime MOR players are completely removed from the population. In the
other regime, the final coexistence of UI and MOR players (for
$0<x_{ui}(0)<0.6$ in BA networks) recovers the partition of ER
networks into Moran-Defectors and Imiting-Cooperators.

The most striking result of the Moran-Imitator competition is found in
the region $0.4<x_{ui}(0)<0.8$ of the BA network. Remarkably, in this
regime the coexistence (which is only a transient for
$0.6<x_{ui}(0)<0.8$) between MOR and UI enhances the average level of
cooperation with respect to the limit in which only imitators are
present, i.e., the combination of strategies is optimal in terms of
the survival of cooperation. As stated above, in the lower half of the
region coexistence is established between two types of agents, MOR
defectors and UI cooperators pointing out that the localization of
defective behavior in Moran players enhance the ability of imitators
to cooperate. On the other hand, the fact that the phenomenon persists
when the coexistence is only temporary (upper half of the region)
makes it clear that survival of MOR defectors in the asymptotic regime
is not a necessary condition for the improving of cooperation. This
phenomenon is reminiscent of the enhancement of cooperation observed
in \cite{roca:2009b} due to errors, but we want to stress that its
mechanism must be different because the case studied in that paper
refers to a situation in which update rules do not co-evolve with the
strategies.

\subsection{Replicator Dynamics versus Unconditional Imitation}

Given that in the previous two scenarios both imitators and
replicators have remarkably outperformed the ability of MOR players to
survive (albeit MOR is only completely suppressed by UI in
heterogeneous networks for same range of parameters), the competition
between replicators and imitators will unveil what is, on average, the
most effective update mechanism.

\begin{figure}[!t]
\epsfig{file=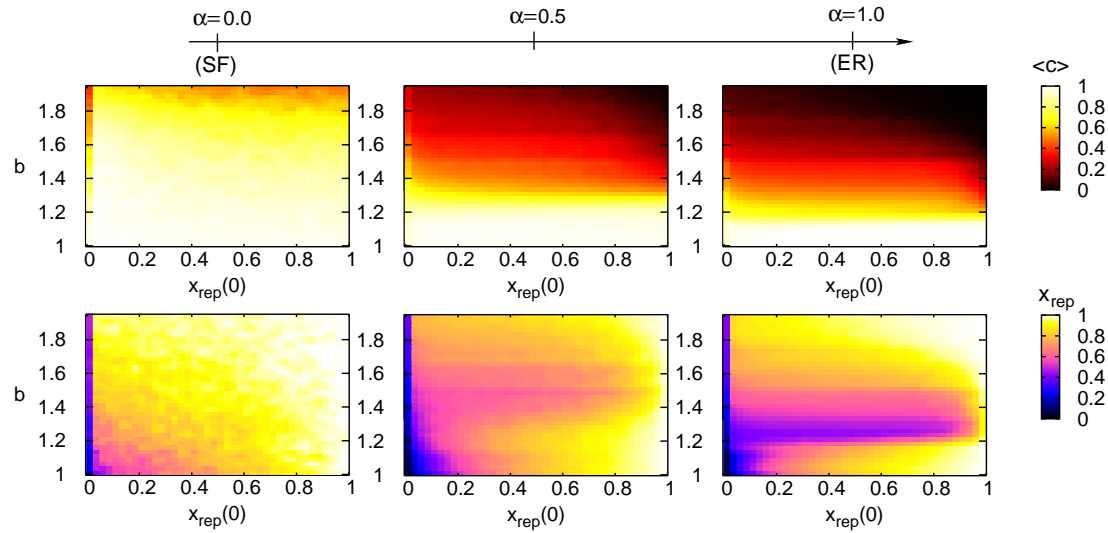,width=0.98\columnwidth,angle=0,clip=1}
\caption{{\em Replicator versus Unconditional Imitation}. The panels
  in the top denote the fraction of cooperators, $\langle c\rangle$,
  in the asymptotic regime in BA networks (left) networks with a
  moderate degree of heterogeneity, denoted as $\alpha=0.5$, (middle)
  and ER graphs (right). The panels in the bottom show the final fraction
  of replicators, $x_{rep}$ for the same cases as above. In all the
  panels both the degree of cooperation and the fraction of
  replicators are plotted as a function of the temptation $b$ and the
  initial fraction of replicators $x_{rep}(0)$.}
\label{repvsui}
\end{figure}

We start by analyzing the BA network. As shown in the bottom-left
panel of Fig. \ref{repvsui}, REP players invade the whole final
population for large values of $b$. However, when the temptation to
defect decreases the final fraction of imitators within the population
increases. In particular, the final fraction of imitators
$x_{ui}=1-x_{rep}$ increases with its initial frequency,
$x_{ui}(0)=1-x_{rep}(0)$. This latter behavior has deep implications
regarding the final degree of cooperation. As shown in the top-left
panel of Fig.~\ref{repvsui}, cooperators tend to survive the more the
larger $x_{ui}(0)$ is (with the exception of the case
$x_{ui}(0)=1$). This result points out that the existence of any small
fraction of imitators coexisting with a majority of replicators
promote cooperation in the system. It is very interesting to realize
that the cooperation level on BA networks is lower when the update
rule is UI than when REP is the rule \cite{roca:2009a} . Therefore,
what we are observing here is that, when both types of dynamics are
mixed, a non-trivial coupling between them arises such that the final
cooperation level is higher when some UI players are left in the
population. Again, this points out to a diversity effect similar to
the one reported in \cite{roca:2009b}, but the same caveat we
expressed above applies.

When the heterogeneity of the network decreases (middle and right
panels) we observe from the panels in the top that, as usual, the
degree of cooperation decreases. However, this decrease in the total
fraction of cooperators is accompanied by the survival of imitators in
some regions of the parameter space as shown by the bottom panels. In
particular, for moderate values of $x_{rep}(0)$ and intermediate
values of $b$ ($b\simeq 1.5$ for $\alpha=0.5$ and $b\simeq 1.3$ in ER
networks) we find a significant fraction of the population playing as
UI. These UI players constitute the main source of cooperation in the
systems since we have checked that the remaining part of the
population, composed of replicator players, are defectors. Therefore,
the natural trend of REP players (as observed from the corresponding
panels in Fig. \ref{morvsrep}) to defect in this region is opposed by
the ability of imitators to cooperate for larger values of
$b$. Outside these two regions, we observe again the prevalence of
replicators which play as cooperators and defectors for small and
large values of $b$ respectively.


\section{Conclusions}
\label{sec:4}

In summary, in this paper we have largely extended the work on
co-evolution of strategies and update rules on square lattices
reported in \cite{moyano:2009}. We have focused on the role of the
degree of heterogeneity of the network by considering a model that
allows to go from BA to ER networks. The main conclusion of this work
is that in all the cases considered, the co-evolution process leads to
the survival of cooperation, even when the temptation to defect is
relatively high.  This allows us to conclude that generically,
co-evolutionary dynamics on networks is very different from the
mean-field problem. In addition, the study we have carried out sheds
more light on the picture that emerged from the preliminary work
\cite{moyano:2009}: While in that paper it was reported that on square
lattices REP displaced UI which in turn displaced MOR, we have learned
here that this statement requires a number of qualifications. Thus, it
is actually the case that REP displaces MOR in all networks studied,
but the other two competitions do not show such a clear-cut
outcome. As we have described in detail above, UI displaces MOR
completely only when the co-evolution process takes place on BA
networks and the temptation parameter is large, whereas a fraction of
UI survives when competing with REP in all the networks studied, at
least for a range of temptation values and initial condition
frequencies.

The asymptotic coexistence of update rules leads to interesting
phenomena such as the complete segregation between UI cooperators and
MOR defectors, or the enhancement of cooperation as compared to the
case when only one rule is present in a number of cases. The first
feature, the segregation phenomenon, is likely to arise due to the
fact that UI cooperators survive through clustering, whereas UI
defectors cannot imitate their MOR neighbors (the rule applies only if
the payoff is strictly larger than the player's) but the converse is
possible. Therefore, compact groups of cooperators are UI, and the
`outside sea' of defection necessarily ends up being MOR. Concerning
the second observation, the enhancement of cooperation by the
co-evolution of update rules, we have already mentioned its
resemblance to the improvement of cooperation by errors reported in
\cite{roca:2009b}. However, it is clear that the relation between the
two mechanisms can only be established through the existence of
different time scales involved in the global dynamics, as in
\cite{roca:2009b} the update rules are fixed. In the preliminary
report \cite{moyano:2009}, the relevance of the time scales was
already noted, but it was also pointed out that the mechanism behind
it is far from clear. Time scales are relevant because rules that are
more prone to change, and therefore lead to more frequent strategy
changes, carry the seed of their own extinction, because in our setup
they will copy the update rules that lead to less frequent
changes. However, the tests carried out in \cite{moyano:2009} with two
versions of UI, one acting only every 10 time steps, showed that while
this was part of the ingredients of the co-evolution, it could not
explain by itself the whole set of observations.  Our work here gives
far more support to the idea that time scales play a key role, by
showing that, even for the same rule, the heterogeneity of the network
induces different rates of invasion of different nodes. All this makes
it clear that it is necessary to design a specific setup in which the
time scales are controlled.

From a more general viewpoint, our work supports the claim that, if
update rules grounded on an evolutionary framework are the ones that
should be used, REP is by far the most likely candidate among the
three ones we have studied. Of course, we do not claim that we have
completely solved this issue. Given the non-triviality of the
co-evolutionary process, a three-rule competition should be analyzed
to confirm this statement, as when the three rules are present the
dynamics could be different. Furthermore, other rules, such as the
Fermi rule \cite{szabo:1998} or the best-response rule
\cite{ellison:1993,roca:2009} should also be included in the
competition. Another important point would be to consider other
games. We can reasonably expect our results to be valid in a parameter
region close to the weak PD, reaching into the strict PD and the
Snowdrift game.  However, we can not predict how far this region
extends and whether games with different equilibrium structure, such
as the Stag-Hunt, would give rise to the same results we report
here. Clearly, all this calls for a more ambitious study for which the
present report is paving the way.  Finally, we find it particularly
interesting that the observation of coexistence in the asymptotic
state of some rules agrees with experimental findings of diversity in
the behavior of humans playing PD games on lattices
\cite{traulsen:2010,grujic:2010}.  Further work is needed to compare
in detail the outcome of the co-evolutionary process with the
available experimental data.

\ack

This work has been partially supported by MICINN (Spain) through
Grants MOSAICO (D.V. and A.S.), FIS2008-01240 and MTM2009-13838
(J.G.G.), by Comunidad de Madrid (Spain) through Grant MODELICO-CM
(A.S.), and by the European Comission through the HPC-EUROPA2 project
228398 (A.C.). D.V.\ acknowledges support from a Postdoctoral Contract
from Universidad Carlos III de Madrid.

\section*{References}

\bibliography{evol-coop}

\end{document}